\documentclass[showpacs,amsmath,amssymb,floatfix]{revtex4}

\usepackage{epsfig}
\usepackage{graphicx}
\usepackage{bm}
\usepackage{amsfonts}

\def\bfig{\begin{figure}[ht] \begin{center}}
\def\bfigh{\begin{figure}[h!] \begin{center}}
\def\bfigb{\begin{figure}[hb!] \begin{center}}
\def\bfigt{\begin{figure}[t!] \begin{center}}
\def\bfight{\begin{figure}[ht!] \begin{center}}
\def\efig{\end{center} \end{figure}}

\def\btab{\begin{table*}[ht]}
\def\etab{\end{table*}}

\newcommand{\vs}{V_\sigma(\sigma)}
\newcommand{\s}{\sigma}

\newcommand{\vp}{V_\phi(\phi)}
\newcommand{\xp}{x_\phi}
\newcommand{\xs}{x_\sigma}
\newcommand{\alp}{\alpha_\phi}
\newcommand{\als}{\alpha_\sigma}

\begin{document}

\title{Quintom evolution in power-law potentials}

\author{Emmanuel N.~Saridakis }
\email{msaridak@phys.uoa.gr} \affiliation{Department of Physics,
University of Athens, GR-15771 Athens, Greece}

\begin{abstract}
We investigate quintom evolution in power-law potentials. We
extract the early-time, tracker solutions, under the assumption of
matter domination. Additionally, we derive  analytical solutions
at intermediate times, that is at low redshifts, which is the
period during the transition from matter to dark-energy
domination. A comparison with exact evolution reveals that the
tracker solutions are valid within $98\%$ for $z\gtrsim1.5$, while
the intermediate ones are accurate within $98\%$ up to
$z\approx0.5$. Using these expressions we extract two new
$w$-parametrizations, one in terms of the redshift and one in
terms of the dark-energy density-parameter, and we present various
quintom evolution sub-classes, including quintessence-like or
phantom-like cases, realization of the $-1$-crossing, and
non-monotonic $w$-evolution.
\end{abstract}

 \maketitle

\section{Introduction}

According to cosmological observations the universe is
experiencing an accelerated expansion, and the transition to the
accelerated phase has been realized in the recent cosmological
past \cite{observ}. In order to explain this remarkable behavior,
one can modify the theory of gravity \cite{ordishov}, or construct
``field'' models of dark energy. The field models that have been
discussed widely in the literature consider a canonical scalar
field (quintessence) \cite{quint0,quint,quint01}, a phantom field,
that is a scalar field with a negative sign of the kinetic term
\cite{phant}, or the combination of quintessence and phantom in a
unified model named quintom \cite{quintom}. Such a combined
consideration intends to describe the crossing of the dark-energy
equation-of-state parameter $w$ through the phantom divide $-1$
\cite{c14}, which is not possible in simple quintessence or simple
phantom scenarios where $w$ lies always on the same side of that
bound ($w>-1$ for quintessence and $w<-1$ for phantom).

Although one can find potential-independent solutions in field
dark energy models \cite{Scherrer:2007pu}, in general the
cosmological evolution depends significantly on the potential
choice. The power-law potential is a widely studied case since,
amongst others, it allows for a theoretical justification through
supersymmetric considerations \cite{Binetruy:1998rz}. For
quintessence \cite{quint0,Watson:2003kk,Kneller:2003xg} or phantom
case \cite{Saridakis:2009pj} it has been shown that it exhibits
``tracker'' behavior, that is for a large class of initial
conditions the cosmological evolution converges to a common
solution at late times \cite{quint01}, which can be extracted
analytically. At intermediate times the dark-energy scalar-field
is still small, but non-negligible, and one has to rely on
perturbative analytical expressions
\cite{Watson:2003kk,Saridakis:2009pj}. However, such an extension
to low redshift is very useful, since observations like supernovae
Ia, WMAP and SDSS ones, are related to this period \cite{observ}.
Finally, an additional advantage of these models is that the field
energy density  remains small at early and intermediate times and
thus the known cosmological epochs are not disturbed.

In this work we desire to study the quintom scenario in power-law
potentials, both at high and low redshifts, and provide the
tracker solutions and the perturbative analytical expressions
respectively. Due to the complex nature of the model, we expect
that the results will be qualitatively different from  simple
quintessence and simple phantom models, even at the tracker level.
The plan of the work is as follows: In section \ref{zeroord} we
construct the quintom paradigm in power-law potentials and we
extract the matter-dominated solutions. In section \ref{firstord}
we derive the cosmological solutions at intermediate times, that
is when dark energy is non-negligible, but still sub-dominant
comparing to the matter content of the universe. In section
\ref{analnum} we compare our analytical expressions with the exact
numerical evolution of the quintom scenario, and in section
\ref{discuss} we provide some applications of the model,
extracting two new $w$-parametrizations and discussing the
cosmological implications. Finally, section
 \ref{concl} is devoted to the summary of our results.

\section{Early time cosmological evolution: tracker solutions} \label{zeroord}

We start by constructing the simple quintom cosmological scenario
in a flat spacetime. The action of a universe constituted of a
canonical $\phi$ and a phantom $\s$  field,  is \cite{quintom}:
\begin{eqnarray}
S=\int d^{4}x \sqrt{-g} \left[\frac{1}{2} R
-\frac{1}{2}g^{\mu\nu}\partial_{\mu}\phi\partial_{\nu}\phi+\vp+
\frac{1}{2}g^{\mu\nu}\partial_{\mu}\sigma\partial_{\nu}\sigma+\vs
+\cal{L}_\text{m}\right], \label{actionquint}
\end{eqnarray}
where we have set $8\pi G=1$.  $\vp$ and $\vs$ are respectively
the quintessence and phantom field potentials, while the term
$\cal{L}_\text{m}$ accounts for the (dark) matter content of the
universe, considered as dust. Finally, although we could
straightforwardly include baryonic matter and radiation in the
model, for simplicity reasons we neglect them since we are
interested in $z<20$ era. The Friedmann equations read
\cite{quintom}:
\begin{equation}\label{FR1}
H^{2}=\frac{1}{3}\Big(\rho_{m}+\rho_{\phi}+\rho_{\s}\Big),
\end{equation}
\begin{equation}\label{FR2}
\dot{H}=-\frac{1}{2}\Big(\rho_{m}+\rho_{\phi}+p_{\phi}+\rho_{\s}+p_{\s}\Big),
\end{equation}
and the evolution equations for the canonical and the phantom
fields are:
\begin{eqnarray}\label{eom}
&&\dot{\rho}_\phi+3H(\rho_\phi+p_\phi)=0,\\
&&\dot{\rho}_\s+3H(\rho_\s+p_\s)=0,
\end{eqnarray}
where $H=\dot{a}/a$ is the Hubble parameter and $a = a(t)$ the
scale factor. In these expressions, $\rho_m$ is the dark matter
density, and $p_\phi$ and $\rho_{\phi}$ are respectively the
pressure and density of the canonical field, while  $p_\s$ and
$\rho_{\s}$ are the corresponding quantities for the phantom
field. They are given by:
\begin{eqnarray}\label{rhophi}
 \rho_{\phi}&=& \frac{1}{2}\dot{\phi}^{2} + V_\phi(\phi)\\
 p_{\phi}&=&  \frac{1}{2}\dot{\phi}^{2} - V_\phi(\phi),\label{pphi}
\end{eqnarray}
\begin{eqnarray}\label{rhosigma}
 \rho_{\s}&=& -\frac{1}{2}\dot{\s}^{2} + V_\s(\s)\\
 p_{\s}&=& - \frac{1}{2}\dot{\s}^{2} - V_\s(\s).\label{psigma}
\end{eqnarray}
Using these expressions, we can equivalently write the  evolution
equation in field terms as:
\begin{eqnarray}
\label{phiddot} &&\ddot{\phi}+3H\dot{\phi}+\frac{\partial
V_\phi(\phi)}{\partial\phi}=0\\
\label{sigmaddot}
 && \ddot{\s}+3H\dot{\s}-\frac{\partial
V_\s(\s)}{\partial\s}=0.
\end{eqnarray}
The equations close by considering the evolution of the matter
density, which in the case of dust reads simply
$\dot{\rho}_m+3H\rho_m=0$, leading to $\rho_m=\rho_{m0}/a^3$, with
$\rho_{m0}$ its value at present.

In quintom scenario, dark energy is attributed to the combination
of $\phi$ and $\s$. Thus, we can define the dark energy density
and pressure as $\rho_{DE}\equiv \rho_{\phi}+ \rho_{\s}$ and
$p_{DE}\equiv p_{\phi}+ p_{\s}$. Therefore, the dark energy
equation-of-state parameter reads
\begin{equation}
\label{wdef}
w\equiv\frac{p_{DE}}{\rho_{DE}}=\frac{p_\phi+p_\s}{\rho_\phi+\rho_\s}.
\end{equation}

In the present work we study quintom evolution in power-law
potentials. That is we consider:
\begin{eqnarray} \label{potphi}
&&\vp = \kappa_\phi\phi^{-\alpha_\phi}\\
&&\vs = \kappa_\s\s^{-\alpha_\s},
 \label{pots}
\end{eqnarray}
where $\kappa_\phi$ and $\kappa_\s$ are constants with units
$m^{4+\alpha_\phi}$ and $m^{4+\alpha_\s}$ respectively. Note that
we do not restrict the values of $\alpha_\phi$ and $\alpha_\s$ a
priori. However, as we are going to see, energy positivity will
force $\alpha_\phi$ to be positive and $\alpha_\s$ to be negative
and bounded, that is potential (\ref{potphi}) will be an inverse
power-law one while (\ref{pots}) will be of a normal power-law
form.

We can express the aforementioned cosmological system using the
scale factor $a$ as the independent variable, since it is
straightforwardly related to the redshift $z$ which is used in
observations. Generalizing \cite{Watson:2003kk,Saridakis:2009pj}
we define
\begin{eqnarray}
 && x_\phi =
{\rho_\phi + p_\phi\over 2(\rho_m + \rho_\phi)} = {{1\over 2}
\dot\phi^{2}\over 3H^2} = {1\over 6}{\Big(a{d\phi\over
da}\Big)}^2, \label{xdefp}\\
 && x_\s=
{\rho_\s + p_\s\over 2(\rho_m + \rho_\s)} = {-{1\over 2}
\dot\s^{2}\over 3H^2} = -{1\over 6}{\Big(a{d\s\over da}\Big)}^2,
\label{xdefp}
\end{eqnarray}
and thus (\ref{rhophi}),(\ref{rhosigma}) can be written as:
 \begin{eqnarray}
 \label{phidotx}
\frac{1}{2} \dot\phi^2 = {x_\phi\over 1-x_\phi}[\rho_m + \vp],\\
-\frac{1}{2} \dot\s^2 = {x_\s\over 1-x_\s}[\rho_m + \vs].
\label{sigmadotx}
\end{eqnarray}
 Therefore, we can
simply write:
\begin{eqnarray}
 &&\rho_{\phi} = \frac{\xp\rho_m + \vp}{1-\xp}\label{rhox}\\
 &&\rho_{\s} = \frac{\xs\rho_m + \vs}{1-\xs}\label{rhoxs}\\
 &&p_{\phi}=\frac{\xp\rho_m - \vp(1-2\xp)}{1-\xp}\label{px}\\
 &&p_{\s}=\frac{\xs\rho_m - \vs(1-2\xs)}{1-\xs}\label{pxs},
\end{eqnarray}
and thus for the dark energy equation-of-state parameter we
obtain:
\begin{equation}
 w = \frac{\frac{\xp\rho_m - \vp(1-2\xp)}{1-\xp}+\frac{\xs\rho_m - \vs(1-2\xs)}{1-\xs}}
 {\frac{\xp\rho_m + \vp}{1-\xp}+\frac{\xs\rho_m + \vs}{1-\xs}}\label{wx}.
\end{equation}
Similarly, the first Friedman equation can be written as
\begin{equation}
3H^2 = \frac{\rho_m + \vp+\vs}{1-\xp-\xs}\label{Hx}.
\end{equation}
 Finally, using
expressions (\ref{rhox})-(\ref{pxs}) and (\ref{Hx}) the field
evolution equations (\ref{phiddot}) and (\ref{sigmaddot}) become
\begin{eqnarray}
a^{2}{\phi}'' + {a{\phi}'\over 2}\left(5-3\xp-3\xs\right) +
{3(1-\xp-\xs)\over \rho_m + \vp+\vs}\Big\{{a{\phi}'[\vp+\vs]\over
2}+ {d\vp\over d\phi}\Big\} = 0 \label{eomxp}
\end{eqnarray}
\begin{eqnarray}
a^{2}{\s}'' + {a{\s}'\over 2}\left(5-3\xp-3\xs\right) +
{3(1-\xp-\xs)\over \rho_m + \vp+\vs}\Big\{{a{\s}'[\vp+\vs]\over
2}- {d\vs\over d\s}\Big\} = 0,  \label{eomxs}
\end{eqnarray}
with prime denoting the derivative with respect to $a$. These
equations are exact and account for the complete dynamics of the
quintom scenario.

In this section we desire to present the tracker solutions, that
is the initial-condition-independent solutions during the
matter-dominated era, i.e at early times. Therefore, we consider
$\rho_\phi,|p_\phi|\ll\rho_m$ and $\rho_\s,|p_\s|\ll\rho_m$, or
equivalently $\xp\ll1$ and $|\xs|\ll1$. Under these
approximations, equations (\ref{eomxp}),(\ref{eomxs}) are
simplified as:
\begin{eqnarray}
&&a^{2}{\phi}''_{(0)} + {5a{\phi}'_{(0)}\over 2} + {3\over
\rho_m}\frac{d\vp}{d\phi}\Big|_{\phi=\phi_{(0)}} = 0 \label{eomxzerop}\\
&&a^{2}{\s}''_{(0)} + {5a{\s}'_{(0)}\over 2} - {3\over
\rho_m}\frac{d\vs}{d\s}\Big|_{\s=\s_{(0)}} = 0. \label{eomxzeros}
 \end{eqnarray}
The zero subscript in parentheses ``(0)'' denotes just these
zeroth-order solutions in terms of $\xp$,$\xs$, or equivalently in
terms of the quintessence and phantom energy density, and must not
be confused later on with the subscript $0$ without parentheses
which stands for the present value of a quantity.

Equation (\ref{eomxzerop}) can be easily solved analytically in
the case of power-law potential (\ref{potphi}). The general
solution for $\alp\neq0$ (which is the trivial case of a constant
potential) is
\begin{equation}
 \phi_{(0)} = C_\phi(\alpha_\phi){a}^{\frac{3}{2 + \alpha_\phi}},
\label{solphizero}
\end{equation}
where the function $C_\phi(\alpha_\phi)$ is related to the
potential parameters through
\begin{equation}
C_\phi(\alpha_\phi) =\Big[\frac{2\alpha_\phi
(2+\alpha_\phi)^2\kappa_\phi}{3\rho_{m0}(4+\alpha_\phi)}\Big]^{\frac{1}{2+\alpha_\phi}}.
 \label{Calphap}
\end{equation}
In (\ref{solphizero}) we have kept only the solution part that
remains small (together with its derivative) for small $a$'s, in
order to be consistent with the matter-dominated  approximation
($\rho_\phi,|p_\phi|\ll\rho_m$). This is the reason for the
absence of initial-condition dependent constants, which is just
the central idea of tracker solutions. In addition, it is easy to
see that this solution remains regular for
$\alpha_\phi\rightarrow-2$. The solution for $\alpha_\phi=-2$ can
be  shown to diverge at small $a$ and thus we do not write it
explicitly.

Similarly, the general solution of equation (\ref{eomxzeros})
under potential (\ref{pots})) is
\begin{equation}
 \s_{(0)} = C_\s(\alpha_\s){a}^{\frac{3}{2 + \alpha_\s}},
\label{solsigmazero}
\end{equation}
where
\begin{equation}
C_\s(\alpha_\s) =\Big[-\frac{2\alpha_\s
(2+\alpha_\s)^2\kappa_\s}{3\rho_{m0}(4+\alpha_\s)}\Big]^{\frac{1}{2+\alpha_\s}}.
 \label{Calphas}
\end{equation}

The zeroth-order solution for $\rho_\phi$ can be calculated from
(\ref{rhox}) under $\xp\ll1$, that is
 $\rho_{\phi (0)} =x_{\phi(0)}\rho_m+V_\phi( \phi_{(0)})$, where $x_{\phi(0)} =
(a\phi'_{(0)})^{2}/6$. The result is:
 \begin{equation}
\label{rhophizero}
 \rho_{\phi (0)} =
3\rho_{m0}{[{C_\phi(\alpha_\phi)]^2}\over {\alpha_\phi(2 +
\alpha_\phi)}}\,a^{-\frac{3\alpha_\phi}{2 + \alpha_\phi}}.
\end{equation}
Furthermore, using the zeroth-order approximation of
(\ref{rhoxs}), i.e $\rho_{\s (0)} =x_{\s(0)}\rho_m+V_\s(
\s_{(0)})$ with $x_{\s(0)} =- (a\s'_{(0)})^{2}/6$, we obtain
 \begin{equation}
\label{rhosigmazero}
 \rho_{\s (0)} =-
3\rho_{m0}{[{C_\s(\alpha_\s)]^2}\over {\alpha_\s(2 +
\alpha_\s)}}\,a^{-\frac{3\alpha_\s}{2 + \alpha_\s}}.
\end{equation}

Similarly to the simple phantom case \cite{Saridakis:2009pj},
expression (\ref{rhosigmazero}) and the requirement of
phantom-energy positivity leads to the constraint
$-2<\alpha_\s<0$, that is the phantom potential is a normal power
law. Additionally, (\ref{rhophizero}) leads to $\alpha_\phi<-2$ or
$\alpha_\phi>0$, with the second case being physically more
interesting since it corresponds to the well-studied inverse
power-law potential of the literature
\cite{quint0,Watson:2003kk,Kneller:2003xg}.

Finally, using the zeroth-order approximation of (\ref{px}) and
(\ref{pxs}), namely
 $p_{\phi (0)} =x_{\phi(0)}\rho_m-V_\phi( \phi_{(0)})$ and
  $p_{\s (0)} =x_{\s(0)}\rho_m-V_\s( \s_{(0)})$ respectively, we
  obtain
 \begin{eqnarray}
\label{pphizero}
 && p_{\phi (0)} = -6\rho_{m0}{{[C_\phi(\alpha_\phi)]^2}\over
{\alpha_\phi(2 +
\alpha_\phi)^2}}\,a^{-\frac{3\alpha_\phi}{2 + \alpha_\phi}}\\
\label{pphizero}
 && p_{\s (0)} =
6\rho_{m0}{{[C_\s(\alpha_\s)]^2}\over {\alpha_\s(2 +
\alpha_\s)^2}}\,a^{-\frac{3\alpha_\s}{2 + \alpha_\s}}.
\end{eqnarray}

At this stage, the zeroth-order, that is early-time, results
coincide with those of simple quintessence \cite{Watson:2003kk} or
simple phantom models \cite{Saridakis:2009pj}. This was expected
since  the exact equations of motion (\ref{eomxp}),(\ref{eomxs}),
which correspond to a coupled system for the two fields, under the
zeroth-order approximation lead to the un-coupled equation system
(\ref{eomxzerop}),(\ref{eomxzeros}). Thus, at early times the two
fields evolve independently. However, even at this approximation
level, the complexity of the quintom model leads to qualitatively
different results. Indeed, dark energy is attributed to both
fields under definition (\ref{wdef}). Therefore, we acquire:
\begin{equation}
\label{wzerosol}
   w_{(0)} =\frac{ p_{\phi (0)}+p_{\s (0)}}{\rho_{\phi (0)}+\rho_{\s (0)}}=
   \frac{-2\frac{[C_\phi(\alpha_\phi)]^2}{\alp(2+\alp)^2}\,a^{-\frac{3\alp}{2+\alp}}+2\frac{[C_\s(\als)]^2}{\als(2+\als)^2}\,a^{-\frac{3\als}{2+\als}}}
   {\frac{[C_\phi(\alpha_\phi)]^2}{\alp(2+\alp)}\,a^{-\frac{3\alp}{2+\alp}}-\frac{[C_\s(\als)]^2}{\als(2+\als)}\,a^{-\frac{3\als}{2+\als}}}.
 \end{equation}
As we observe, although in the total absence of one of the fields
this expression gives the corresponding results of  simple
quintessence ($w_{(0)}=-{2\over
   2+\alp}$ \cite{Watson:2003kk}) and simple phantom ($w_{(0)}=-{2\over
   2+\als}$ \cite{Saridakis:2009pj}), when both fields are present
   and despite the fact that they are not coupled at this zeroth-level, the
   behavior of $w_{(0)}$ is qualitative different. In particular, as can
   be clearly seen, $w_{(0)}$ is not a constant but it is
   $a$-dependent although it is the tracker solution. On the
   contrary, in simple quintessence and simple phantom scenarios
   one should go beyond tracker solutions, that is to intermediate
   times, in order to reveal a varying equation-of-state
   parameter. This feature makes the model at hand cosmologically
   interesting.
We mention that according to the specific model and choice of
parameters, $w_{(0)}$ can evolve below or above $-1$, or even
experience the phantom-divide crossing.

 Finally, we can calculate the
 zeroth-order behavior of the dark-energy density
parameter as:
\begin{equation}
\Omega_{DE(0)}=\frac{\rho_{\phi (0)}+\rho_{\s (0)}}{\rho_{m}}=
3{[{C_\phi(\alpha_\phi)]^2}\over {\alpha_\phi(2 +
\alpha_\phi)}}\,a^{\frac{6}{2 + \alpha_\phi}}-
3{[{C_\s(\alpha_\s)]^2}\over {\alpha_\s(2 +
\alpha_\s)}}\,a^{\frac{6}{2 + \alpha_\s}}
 \label{Om0tilde},
\end{equation}
where we have used that $\rho_m=\rho_{m0}/a^3$.

Expressions (\ref{solphizero})-(\ref{Om0tilde}) are the tracker
solutions for quintom cosmology in power-law potentials.
Equivalently they can be expressed as a function of time,
considering $a\propto t^{2/3}$
 since we are in the matter-dominated era, or as a function of the
redshift $z$ through $\frac{a_0}{a}=1+z$, with $a_0=1$ the present
value. They provide an excellent approximation to the behavior of
the quintom scenario as long as $\rho_\phi,|p_\phi|\ll\rho_m$ and
$\rho_\s,|p_\s|\ll\rho_m$, that is at early times.

\section{Cosmological evolution at intermediate times} \label{firstord}

As time passes and cosmological evolution continues, the scalar
fields increase and dark-energy density becomes non-negligible,
although still dominated by the dark-matter one. Therefore, we
expect that progressively the various quantities will start
diverging from the expressions obtained above. We are interested
in the analytical investigation of this intermediate evolution
stage,  and thus we perform a first-order perturbation to the
zeroth-order solutions of section \ref{zeroord}. The obtained
solutions will be significant for the period during the transition
from matter to dark-energy domination. The corresponding
quantities are denoting by the subscript ``$(1)$'', and the total
ones by tilde, i.e:
\begin{equation}
\label{firstorder}
 \widetilde\phi = \phi_{(0)} + \phi_{(1)}, \  \widetilde\s = \s_{(0)} + \s_{(1)}, \  \widetilde\rho_\phi = \rho_{\phi{(0)}} +
 \rho_{\phi{(1)}},\  \widetilde\rho_\s = \rho_{\s{(0)}} +
 \rho_{\s{(1)}},\  \widetilde w = w_{(0)} + w_{(1)}.
 \end{equation}
The aforementioned perturbations can be equivalently considered as
keeping terms up to first order in $\xp$ and $\xs$ (i.e keeping
only $x_{\phi(0)}$ and $x_{\s(0)}$) and in $\phi_{(1)}/\phi_{(0)}$
and $\s_{(1)}/\s_{(0)}$, in the exact evolution equations
(\ref{eomxp}),(\ref{eomxs}). An additional convenient formula can
arise form the expansions of the potentials as
$V_\phi(\widetilde\phi)=V_\phi(\phi_{(0)})+\frac{d
V_\phi(\phi_{(0)})}{d \phi_{(0)}} \phi_{(1)}+{\mathcal{O}}(
\phi_{(1)}^2)$, and similarly for $V_\s(\widetilde\s)$.
Substituting the expansions (\ref{firstorder}) into
(\ref{eomxp}),(\ref{eomxs}), we acquire the approximate evolution
equations at this intermediate-time region:
\begin{equation}
 a^{2}{\phi_{(1)}''} + {a\over
2}\left\{5\phi_{(1)}'-3[x_{\phi(0)}+x_{\s(0)}]\phi_{(0)}'\right\}
+ {3a\phi_{(0)}'[\vp+\vs]\over 2 \rho_{m}} -{3[\rho_{\phi
(0)}+\rho_{\s (0)}]\over {\rho_{m}}^{2}} {dV_\phi(\phi_{(0)})\over
d\phi_{(0)}}+ {3 \over \rho_{m}}{d^2 V_\phi(\phi_{(0)}) \over
d\phi_{(0)}^2}\phi_{(1)} = 0, \label{eomxfristordp}
 \end{equation}
 \begin{equation}
 a^{2}{\s_{(1)}''} + {a\over
2}\left\{5\s_{(1)}'-3[x_{\phi(0)}+x_{\s(0)}]\s_{(0)}'\right\} +
{3a\s_{(0)}'[\vp+\vs]\over 2 \rho_{m}} +{3[\rho_{\phi
(0)}+\rho_{\s (0)}]\over {\rho_{m}}^{2}} {dV_\s(\s_{(0)})\over
d\s_{(0)}}- {3 \over \rho_{m}}{d^2 V_\s(\s_{(0)}) \over
d\s_{(0)}^2}\s_{(1)} = 0, \label{eomxfristords}
 \end{equation}
  where $x_{\phi(0)} =
(a\phi'_{(0)})^{2}/6$ and $x_{\s(0)} =- (a\s'_{(0)})^{2}/6$.
Inserting the explicit power-law potential forms (\ref{potphi})
and (\ref{pots}) we finally obtain
 \begin{eqnarray}
a^{2}{\phi_{(1)}''} + {a\over
2}\left[5\phi_{(1)}'-\frac{a^2}{2}\,\phi_{(0)}'^3\right]
+\frac{a^3}{4}\,\s_{(0)}'^2\phi_{(0)}'+ {27 \kappa_\phi a^{3}
\phi_{(0)}^{1-\alp}\over 2(2+\alp)\rho_{m0}}
  +
  \frac{3\left[\kappa_\s\s_{(0)}^{-\als}-\frac{\rho_{m0}}{6a}\s_{(0)}'^2\right]a^6}{\rho_{m0}^2}\kappa_\phi\alp\phi_{(0)}^{-1-\alp}+\nonumber\\
  +
  \frac{3a^4}{2\rho_{m0}}\kappa_\s\s_{(0)}^{-\als}\phi_{(0)}'+
{3\alp(1+\alp)a^{3}\kappa_\phi \phi_{(0)}^{-2-\alp}\phi_{(1)}\over
\rho_{m0}} = 0 \label{eomxfristordp2}
\end{eqnarray}
and
 \begin{eqnarray}
a^{2}{\s_{(1)}''} + {a\over
2}\left[5\s_{(1)}'+\frac{a^2}{2}\,\s_{(0)}'^3\right]
-\frac{a^3}{4}\,\phi_{(0)}'^2\s_{(0)}'+ {27 \kappa_\s a^{3}
\s_{(0)}^{1-\als}\over 2(2+\als)\rho_{m0}}
  -
  \frac{3\left[\kappa_\phi\phi_{(0)}^{-\alp}+\frac{\rho_{m0}}{6a}\phi_{(0)}'^2\right]a^6}{\rho_{m0}^2}\kappa_\s\als\s_{(0)}^{-1-\als}+\nonumber\\
  +
  \frac{3a^4}{2\rho_{m0}}\kappa_\phi\phi_{(0)}^{-\alp}\s_{(0)}'-
{3\als(1+\als)a^{3}\kappa_\s \s_{(0)}^{-2-\als}\s_{(1)}\over
\rho_{m0}} = 0. \label{eomxfristords2}
\end{eqnarray}
Despite their complicated form, these equation allow for a
relatively simple general solution, if we keep only the part that
remains small (together with its derivative) for small $a$'s, in
order to be consistent with the matter-dominated  approximation.
For $\alp>0$ and $-2<\als<0$ it reads:
\begin{eqnarray}
&& \phi_{(1)} =   B_\phi(\alp,\als)\,{a}^{\frac{9}{2 +
\alpha_\phi}},
\label{phifirstorder}\\
&& \s_{(1)} =   B_\s(\alp,\als)\,{a}^{\frac{9}{2 + \alpha_\s}},
\label{sigmafirstorder}
\end{eqnarray}
where the functions $B_\phi(\alp,\als)$ and $B_\s(\alp,\als)$ are
related to the potential parameters through
\begin{eqnarray}
&&B_\phi(\alp,\als)
=3C_\phi(\alp)\frac{\{-[C_\phi(\alp)]^2\als(2+\als)^2(6+\alp)+[C_\s(\als)]^2\alp(2+\alp)[12+4(\alp+\als)+\alp\als]\}}
  {\alp(2+\alp)\als(2+\als)^2(28+8\alp+\alp^2)}
 \label{Bphap}\\
&&B_\s(\alp,\als)
=-3C_\s(\als)\frac{\{-[C_\s(\als)]^2\alp(2+\alp)^2(6+\als)+[C_\phi(\alp)]^2\als(2+\als)[12+4(\alp+\als)+\alp\als]\}}
  {\als(2+\als)\alp(2+\alp)^2(28+8\als+\als^2)},\ \ \ \ \ \ \
 \label{Bphaps}
\end{eqnarray}
(that is $B_\s(\alp,\als)$ arises form $B_\phi(\alp,\als)$
changing the index $\phi$ to $\s$ and vice versa, and adding an
overall minus sign). In these expressions the complexity of the
quintom model becomes clear, since they are qualitatively
different from those of simple quintessence \cite{Watson:2003kk}
or simple phantom \cite{Saridakis:2009pj} models. Indeed, since
the first-order approximated evolution equations
(\ref{eomxfristordp2}),(\ref{eomxfristords2}) are coupled, their
general solution reveals a strong interdependence of the two
fields, too. This feature will become even stronger in the
$w_{(1)}$-solution below, which by definition it already depends
on both fields. Finally, in the total absence of one field,
expressions (\ref{Bphap}), (\ref{Bphaps}) give exactly those of
the corresponding simple models
\cite{Watson:2003kk,Saridakis:2009pj}.

The  perturbation of $\rho_\phi$ can be calculated from
(\ref{rhox}) keeping the corresponding terms, thus:
\begin{equation}
\rho_{\phi (1)} = x_{\phi(0)}\rho_{\phi (0)} + x_{\phi(1)}\rho_m -
\alp{\phi_{(1)}\over \phi_{(0)}}\vp,\label{rhophifirstorder}
\end{equation}
where $x_{\phi(1)} = \frac{1}{3}a^2\phi_{(0)}'\phi_{(1)}'$ as it
arises from $\widetilde{x}_\phi = x_{\phi(0)} + x_{\phi(1)}=
a^{2}(\phi'^{2}_{(0)} + 2\phi'_{(0)}\phi'_{(1)})/6$. Therefore,
using also (\ref{solphizero}), (\ref{rhophizero}) and
(\ref{phifirstorder}) in order to express the result in terms of
$\rho_{\phi(0)}$, we obtain
\begin{equation}
 \rho_{\phi (1)} = \Big\{ \frac{3[C_\phi(\alp)]^3-B_\phi(\alp,\als)\alp(\alp^2-4)}{2C_\phi(\alp)(2+\alp)^2}\Big\}\rho_{\phi
 (0)}\,a^{\frac{6}{2+\alp}}.\ \
\label{rhophifirstordersolna}
 \end{equation}
Repeating the same procedure for $\rho_\s$ we acquire:
\begin{equation}
 \rho_{\s (1)} = -\Big\{ \frac{3[C_\s(\als)]^3+B_\s(\alp,\als)\als(\als^2-4)}{2C_\s(\als)(2+\als)^2}\Big\}\rho_{\s (0)}\,a^{\frac{6}{2+\als}}.
\label{rhophifirstordersolnas}
 \end{equation}

The  perturbation of $w$ can arise form (\ref{wdef}) as
\begin{equation}
w_{(1)} =\frac{p_{\phi (1)}+p_{\s (1)}}{\rho_{\phi (0)}+\rho_{\s
(0)}}- \frac{p_{\phi (0)}+p_{\s (0)}}{[\rho_{\phi (0)}+\rho_{\s
(0)}]^2}\,[ \rho_{\phi (1)}+\rho_{\s (1)}], \label{wfirstorder0}
\end{equation}
with $p_{\phi (1)} = x_{\phi(0)}\rho_{\phi (0)} +
x_{\phi(1)}\rho_m + \alp{\phi_{(1)}\over \phi_{(0)}}\vp$ (and
similarly for $p_{\s(1)}$). The final result reads
\begin{equation}
w_{(1)}
=\frac{\alp^2(2+\alp^2)\als^2(2+\als^2)a^{\frac{12(\alp+\als+\alp\als)}{(2+\alp)(2+\als)}}}
{2\left\{[C_\s(\als)]^2
\alp(2+\alp)a^{\frac{3\alp}{2+\alp}}-[C_\phi(\alp)]^2
\als(2+\als)a^{\frac{3\als}{2+\als}}\right\}^2}\,(d_1+d_2+d_3+d_4),
\label{wfirstorder}
\end{equation}
where
\begin{eqnarray}
&&d_1=\frac{[C_\phi(\alp)]^3(4+\alp)\{3[C_\phi(\alp)]^3+B_\phi(\alp,\als)\alp(2+\alp)(6+\alp)\}}
{\alp^2(2+\alp)^5}a^{\frac{6(1-\alp)}{2+\alp}}
\nonumber\\
&&d_2=\frac{[C_\s(\als)]^3(4+\als)\{-3[C_\s(\als)]^3+B_\s(\alp,\als)\als(2+\als)(6+\als)\}}
{\als^2(2+\als)^5}a^{\frac{6(1-\als)}{2+\als}}
\nonumber\\
 &&d_3=-\frac{C_\phi(\alp)[C_\s(\als)]^2\{3(4+\als)[C_\phi(\alp)]^3+B_\phi(\alp,\als)\alp(2+\alp)[24+(10+\alp)\als]\}}
 {\alp(2+\alp)^3\als(2+\als)^2}\,a^{\frac{6}{2+\als}-\frac{6\alp}{2+\alp}}
\nonumber\\
  &&d_4=-\frac{C_\s(\als)[C_\phi(\alp)]^2\{-3(4+\alp)[C_\s(\als)]^3+B_\s(\alp,\als)\als(2+\als)[24+(10+\als)\alp]\}}
 {\als(2+\als)^3\alp(2+\alp)^2}\,a^{\frac{6}{2+\alp}-\frac{6\als}{2+\als}}
\nonumber.
\end{eqnarray}
Note that  expressions (\ref{rhophifirstordersolna}),
(\ref{rhophifirstordersolnas}) and (\ref{wfirstorder}) in the
total absence of one field, coincide with those of the
corresponding simple quintessence \cite{Watson:2003kk} and simple
phantom \cite{Saridakis:2009pj} models. The qualitative difference
of quintom scenario is also reflected in the un-specified sign of
the correction terms, contrary to simple models where these
corrections have a constant sign ($\phi_{(1)}$ is always negative
in both quintessence and phantom cases, and $\rho_{\phi(1)}>0$ and
$w_{\phi(1)}<0$  in simple quintessence while $\rho_{\phi(1)}<0$
and $w_{\phi(1)}>0$  in simple phantom scenario).

In summary, inserting  (\ref{phifirstorder}),
(\ref{sigmafirstorder}), (\ref{rhophifirstordersolna}),
(\ref{rhophifirstordersolnas}) and (\ref{wfirstorder}) in the
perturbative expansion (\ref{firstorder}) we finally obtain the
analytical solutions for intermediate times, namely
$\widetilde\phi, \widetilde\s, \widetilde\rho_\phi,
\widetilde\rho_\s$ and $\widetilde w$.
 Finally,
$\widetilde{\Omega}_{DE}$ can be calculated as
 $\widetilde{\Omega}_{DE} =
(\widetilde{\rho}_{\phi}+\widetilde{\rho}_{\s})/(\widetilde{\rho}_{\phi}
+\widetilde{\rho}_{\s}+ \rho_{m})$. We mention  that since we have
obtained the aforementioned quantities as a function of the scale
factor, it is straightforward to express them as a function of the
redshift $z$ through $\frac{1}{a}=1+z$.

\section{Comparing analytical and numerical results}\label{analnum}

In the previous two sections we extracted analytical expressions
for the quintom scenario in general power-law potentials, when the
universe is dominated by the dark matter sector. Concerning their
applicability, we argued that these solutions are valid at early
and intermediate times, that is at both high and low redshift. In
order to examine the precision of our results, we evolve
numerically the exact cosmological system calculating the exact
quantities $\phi(a)$, $\Omega_{DE}(a)$, $w(a)$, and then we study
their divergence from the zeroth and first order formulae derived
above. This can be achieved by examining the ratios
$\phi_{(0)}/\phi$, $\widetilde\phi/\phi$, $\s_{(0)}/\s$,
$\widetilde\s/\s$ and similarly $\Omega_{DE(0)}/\Omega_{DE}$,
$\widetilde\Omega_{DE}/\Omega_{DE}$ and $w_{(0)}/w$, $\widetilde
w/w$. The initial conditions are chosen consistently with initial
matter domination, since if this requirement is fulfilled then the
results do not depend on the specific initial conditions, as
expected. Finally, we fix $\kappa_\phi$ and $\kappa_\s$ in order
to acquire $\Omega_{m0} \approx 0.28$ and $\Omega_{DE0} \approx
0.72$ at present. Note that contrary to simple quintessence and
simple phantom cases, one can satisfy these observational
requirements by many different realizations of the quintom
scenario (i.e many different potential-parameter choices), which
is an additional advantage of the model at hand.

\begin{figure}[ht]
\begin{center}
\mbox{\epsfig{figure=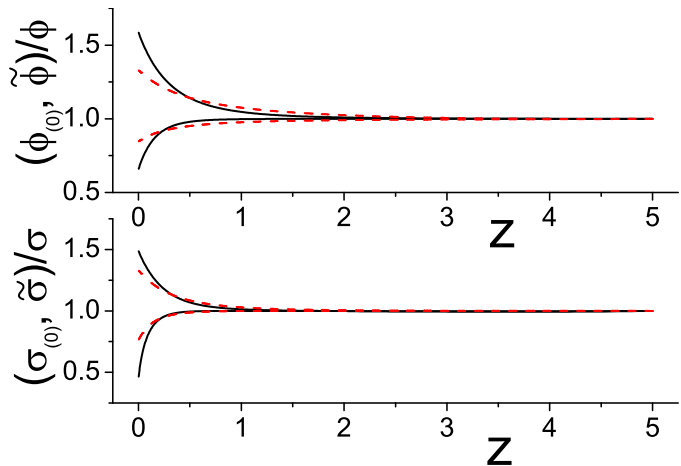,width=9cm,angle=0}} \caption{ (Color
Online){\it Upper graph: The ratios $\phi_{(0)}/\phi$ (upper two
curves) and $\widetilde \phi/\phi$ (lower two curves) as a
function of the redshift, for $\alp = 2,\als=-1$ (black, solid)
and $\alp = 1,\als=-0.5$ (red, dashed). Lower graph: The ratios
$\s_{(0)}/\s$ (upper two curves) and $\widetilde \s/\s$ (lower two
curves) as a function of the redshift, for $\alp = 2,\als=-1$
(black, solid) and $\alp = 1,\als=-0.5$ (red, dashed).}}
\label{fig1phi}
\end{center}
\end{figure}

In fig. \ref{fig1phi} we present the ratios of the zeroth and
first order field solutions to the exact numerical values, as a
function of the redshift.
 The
calculations have been performed for two potential combinations,
namely $\alp = 2,\als=-1$ and $\alp = 1,\als=-0.5$. As we observe,
$\phi_{(0)}/\phi$ and $\s_{(0)}/\s$ are very close to $1$  for
$z\gtrsim1.5$, that is the tracker solution is a very good
approximation at this early evolution stage, with errors of less
than $3\%$. Furthermore, for $1.5\gtrsim z\gtrsim0.5$ the
zeroth-order solution is not a good approximation (with error
$15\%$) but the first-order one remains within $95\%$ accuracy. As
expected these errors are larger than the corresponding results of
simple quintessence \cite{Watson:2003kk} and simple phantom
\cite{Saridakis:2009pj} models, since the exact system is coupled
and thus the evolution depends on both fields. Finally, at late
times, that is for $z<0.5$, the scalar fields are enhanced
significantly, dark energy dominates the universe, and our
approximation breaks down rapidly.

\begin{figure}[ht]
\begin{center}
\mbox{\epsfig{figure=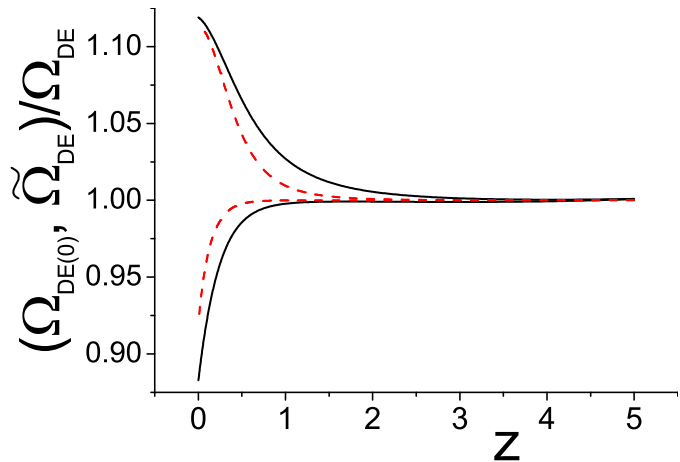,width=9cm,angle=0}} \caption{ (Color
Online){\it The ratios $\Omega_{DE(0)}/\Omega_{DE}$ (upper two
curves) and $\widetilde\Omega_{DE}/\Omega_{DE}$ (lower two curves)
as a function of the redshift, for $\alp = 2,\als=-1$ (black,
solid) and $\alp = 1,\als=-0.5$ (red, dashed).}} \label{fig2om}
\end{center}
\end{figure}

In fig. \ref{fig2om} we depict $\Omega_{DE(0)}/\Omega_{DE}$ and
$\widetilde\Omega_{DE}/\Omega_{DE}$ as a function of $z$.
The divergence of the zeroth-order solution from the exact one
starts at $z\approx1.5$. However, we can see that the first-order
solution is very accurate (with error less than $2\%$) up to
$z\approx0.5$. After that point, the first-order solution starts
diverging rapidly from the exact one and our approximation is not
valid.

\begin{figure}[ht]
\begin{center}
\mbox{\epsfig{figure=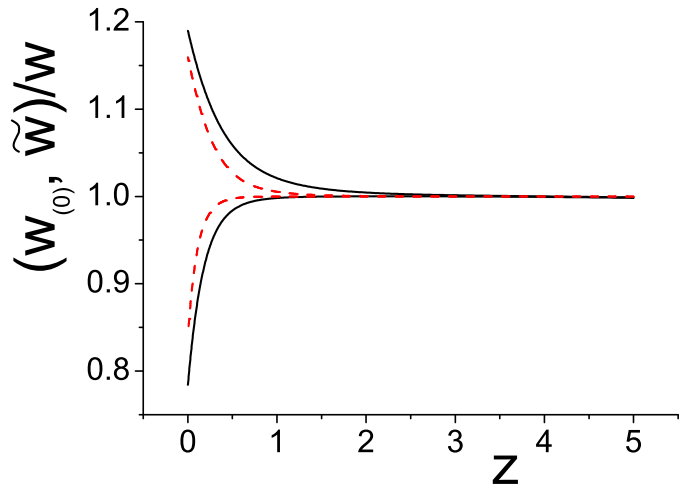,width=9cm,angle=0}} \caption{ (Color
Online){\it The ratios $w_{(0)}/w$ (upper two curves) and
$\widetilde w/w$ (lower two curves) as a function of the redshift,
for $\alp = 2,\als=-1$ (black, solid) and $\alp = 1,\als=-0.5$
(red, dashed).}} \label{fig3w}
\end{center}
\end{figure}

Finally, in fig. \ref{fig3w}  we present $w_{(0)}/w$ and
$\widetilde w/w$ as a function of the redshift.
We observe that the tracker solution is very accurate up to
$z\approx1.5$, while the first-order one agrees with the exact
evolution within $2\%$ up to $z\approx 0.5$. As expected these
errors are larger than the corresponding results of simple models
\cite{Watson:2003kk,Saridakis:2009pj}, since the error of $w$
depends on the errors of both fields.

Therefore, from these tree figures we conclude that the tracker,
that is the zeroth-order, solution is accurate within an error of
$2\%$ at early times and up to $z\approx1.5$. At intermediate
times, that is at low redshifts, we have to use the first-order
analytical solution, which agrees with the exact cosmological
evolution  up to $z\approx0.5$, within an error of $2\%$. After
that point, the dark energy sector enhances, it dominates the dark
matter one, and our approximation breaks down rapidly.

\section{Cosmological implications and discussion} \label{discuss}

Having tested the accuracy of our analytical expressions and
having determined their applicability region, we can use them to
describe an arbitrary quintom evolution in power-law potentials up
to $z=0.5$. Clearly, the tracker solutions are much simpler and
can be used as a first approximation. Contrary to  simple
quintessence and simple phantom models, $w$ presents a varying
behavior even at this zeroth-order approximation level, and thus
it can offer a reliable description of the system (quantitatively
up to $z\approx1.5$ and qualitatively later on). For intermediate
times one can use at will the analytical expressions for
$\widetilde w$ and $\widetilde\Omega_{DE}$, which are very
accurate up to $z\approx0.5$. Here we present an additional
simplification which allows for an easier application. Observing
that their complicated form is due to the $a$-independent
coefficients (which are determined by the potential parameters)
rather than the exponents of $a$, we can make an easy shortening
keeping the most significant $a$-terms and thus resulting to a
convenient simulation of $w$-evolution. Therefore, expressing the
formula in terms of the redshift  $z$ through $\frac{1}{a}=1+z$,
we finally acquire:
\begin{equation}
 w_{fit\text{\,z}}(z)=w_1(1+z)^{s_1}+(w_0-w_1)(1+z)^{s_2}.
\label{wfit}
 \end{equation}
In this $w$-parametrization $w_0$ is its present value, which can
be taken from observations, while $w_1$, $s_1$ and $s_2$ are
determined by the potential exponents in a rather complicated way.
Equivalently, we can consider $w_1$, $s_1$ and $s_2$ as
additional, free parameters. In this way, the relative large
number of parameters in the fit (\ref{wfit}) is attributed to the
general behavior of quintom scenario, which must be able to
present a large class of cosmological evolutions. Furthermore, the
use of three or four parameters in $w$-parametrizations is usual
in the literature \cite{Szydlowski:2006ay}.

\begin{figure}[ht]
\begin{center}
\mbox{\epsfig{figure=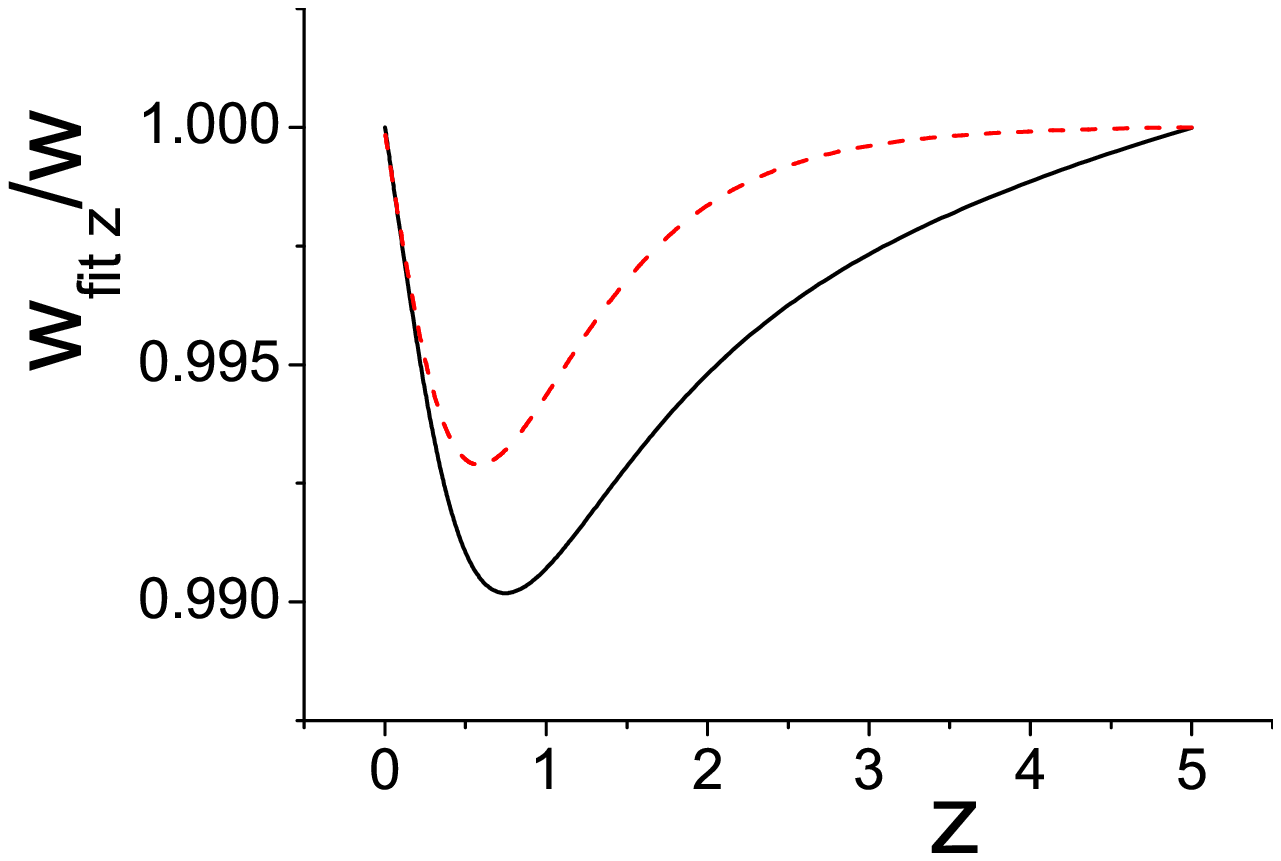,width=9cm,angle=0}} \caption{ (Color
Online){\it The ratio $w_{fit\text{\,z}}/w$, where
$w_{fit\text{\,z}}$ is given by (\ref{wfit}), as a function of
$z$, for $\alp = 2,\als=-1$ (black, solid) and $\alp =
1,\als=-0.5$ (red, dashed).}} \label{fig4wfit}
\end{center}
\end{figure}

Let us examine the accuracy of the fit (\ref{wfit}), comparing it
with the exact cosmological evolution of the model at hand. In
fig. \ref{fig4wfit} we depict the ratio $ w_{fit\text{\,z}}/w$ as
a function of $z$ for two potential choices. In this figure the
value $w_0$ has been determined numerically, that is why
$w_{fit\text{\,z}}$ and $w$ coincide both initially and finally.
As we observe, the proposed $w$-parametrization, given by
(\ref{wfit}), agrees with the exact evolution within $99\%$.
Similarly to the figures of the previous section, the errors are
larger for larger $\alp$ and $|\als|$, however small $\alp$,
$|\als|$ are expected to be more realistic. The very small error
is due to the general and multi-parametric form of the fit
(\ref{wfit}), which allows for a very satisfactory fit of a
general quintom evolution in power-law potentials.

\begin{figure}[ht]
\begin{center}
\mbox{\epsfig{figure=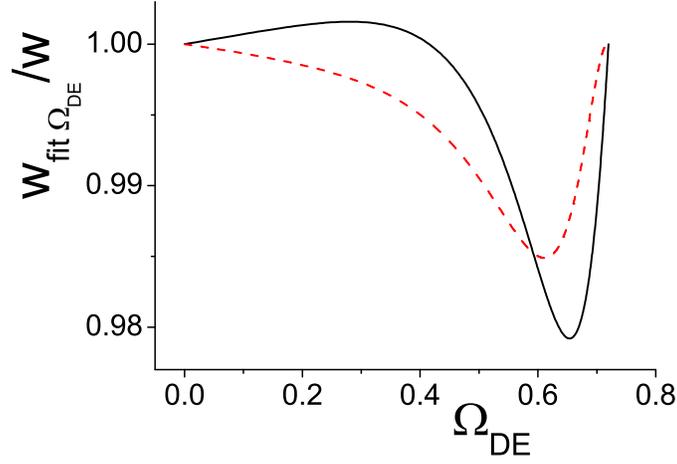,width=9cm,angle=0}} \caption{ (Color
Online){\it The ratio $w_{fit \Omega_{DE}}/w$, where $w_{fit
\Omega_{DE}}$ is given by (\ref{wOmfit}), as a function of
$\Omega_{DE}$, for $\alp = 2,\als=-1$ (black, solid) and $\alp =
1,\als=-0.5$ (red, dashed).}} \label{fig5wOm}
\end{center}
\end{figure}

Another useful approximated relation, that can arise from our
analytical results, is the expression of $w$ as a function of
$\Omega_{DE}$. Indeed, since we have derived the relations $w(a)$
and $\Omega_{DE}(a)$, we can keep the most significant $a$-terms
and then eliminate $a$, resulting to $w(\Omega_{DE})$. Doing so we
obtain:
\begin{equation}
w_{fit \Omega_{DE}}(\Omega_{DE})=w_2+w_3
\Omega_{DE}^{s_3}+w_4\Omega_{DE}^{s_4}. \label{wOmfit}
 \end{equation}
 The various parameters can be determined by the specific potential
 choice in a rather complicated way. Equivalently we can
 consider them as free parameters, suitably fitted in order to
 acquire the best agreement with the exact cosmological evolution.
 Using a $\chi^2$ minimization routine to determine the best-fit
 values of these parameters, in
fig. \ref{fig5wOm} we present the ratio $ w_{fit \Omega_{DE}}/w$
as a function of $\Omega_{DE}$ for two potential choices, where
the present $\Omega_{DE}$-value has been taken to be $0.72$, both
in $ w_{fit \Omega_{DE}}$ and $w$.
The error introduced by the fit (\ref{wOmfit}) is less than
$2.5\%$ for the examined potential cases, mainly at late times.
The fact that it is larger than the corresponding error of
$w_{fit\text{\,z}}$, is a result of the simplifications we
performed in order to eliminate $a$ and obtain $w_{fit
\Omega_{DE}}(\Omega_{DE})$. Finally, we stress that fits
(\ref{wfit}) and (\ref{wOmfit}) can be considered to be valid up
to present epoch ($z=0$ and $\Omega_{DE}\approx0.72$) although
they have arisen from expressions valid up to $z\approx0.5$, since
the relative large number of free parameters leads to small errors
and one can always adjust the fit to coincide with the exact
evolution both initially and finally.

\begin{figure}[ht]
\begin{center}
\mbox{\epsfig{figure=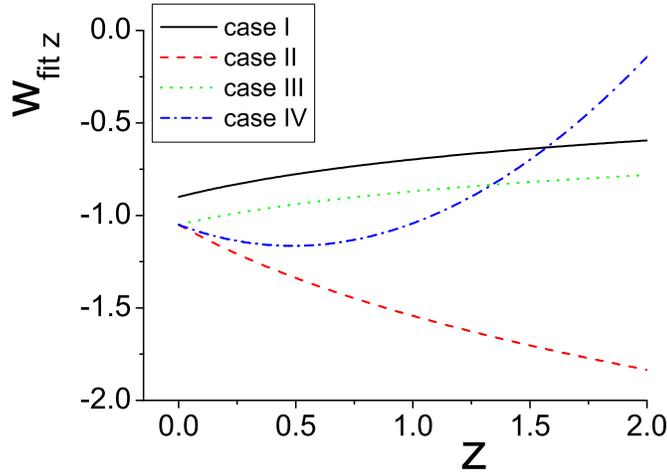,width=9cm,angle=0}} \caption{ (Color
Online){\it $w_{fit\text{\,z}}(z)$ for four different cosmological
cases, as it arises from parametrization (\ref{wfit}). Case I:
$w_0=-0.9$, $w_1=-1$, $s_1=-1/3$, $s_2=1/5$ (black, solid), case
II:
 $w_0=-1.05$, $w_1=1$, $s_1=-1/3$, $s_2=1/5$ (red, dashed), case III:
 $w_0=-1.05$, $w_1=-1$, $s_1=-1/3$, $s_2=1/5$ (green, dotted), case IV:
 $w_0=-1.05$, $w_1=-1.8$, $s_1=1$, $s_2=1.8$ (blue, dashed-dotted).}} \label{fig6examples}
\end{center}
\end{figure}

Since we have examined the accuracy of our $w(z)$ and
$w(\Omega_{DE})$ parametrizations, we can use them at will to
describe a large class of cosmological evolutions. In
fig.~\ref{fig6examples} we depict $w_{fit\text{\,z}}(z)$ for four
different cosmological cases.
 Case I corresponds to a quintessence-like evolution,
where $w$ remains always larger than $-1$. Case II corresponds to
a phantom-like evolution, with $w<-1$. Case III corresponds to a
quintom evolution where $w$ crosses the phantom divide in the
recent cosmological past, as might be suggested by observations
within $95\%$ confidence level \cite{observ}. It is interesting to
see that (\ref{wfit}) allows for a non-monotonic
$w_{fit\text{\,z}}(z)$, too. Indeed, case IV corresponds to such a
behavior, where $w$ experiences the $-1$-crossing from above to
below and then it starts increasing again.

In summary, as was mentioned above, expression (\ref{wfit}) can
describe a very large class of cosmological evolutions. Similar
results can be obtained using $w_{fit \Omega_{DE}}(\Omega_{DE})$
from (\ref{wOmfit}), where one can acquire various
$w(\Omega_{DE})$ cosmological behaviors. Finally, it is easy to
see, that the case of simple quintessence can be described by
(\ref{wfit}) and (\ref{wOmfit}) using $s_1=0$,
$s_2=-\frac{6}{2+\alp}$, $s_3=1$, $w_1=w_2=-\frac{2}{2+\alp}$,
$w_3=- {2 \alp (4 + \alp) \over (2+\alp)(\alp^2+8\alp+28)}$,
$w_4=0$, while for the simple phantom model the corresponding
parameters are the same with $\alp\rightarrow\als$. However, it is
more convenient to consider them as free parameters.

\section{Conclusions}
\label{concl}

In this work we studied quintom models with power-law potentials.
Firstly, we extracted the tracker solutions under the assumption
of matter domination. These solutions correspond to the general
and common behavior of all such models at early times, that is at
high redshifts. Contrary to simple quintessence and simple phantom
cases, even at this zeroth-order approximation level, $w$ is not
constant but evolving, which is a result of the complex nature of
quintom scenario. The tracker solutions are quantitatively valid
up to $z\approx1.5$ (within $98\%$ accuracy), but since this
zeroth-order approximation possesses the qualitative features of
the exact solution (varying and not constant $w$), it could be
used up to low redshifts in order to provide a first picture of
the corresponding cosmological evolution.

In addition, we extracted the general cosmological solutions at
intermediate times, that is at low redshifts, which is the period
during the transition from matter to dark-energy domination. Such
a solution can be very useful in dark energy observations, since
probes based on supernovae Ia, WMAP and SDSS,  are related to this
cosmological epoch \cite{observ}. The comparison with the exact
evolution shows that these first-order solutions are accurate
within $2\%$ up to $z\approx0.5$.

In order to use the aforementioned results in a relative simple
way, we have further approximated them extracting two new
$w$-parametrizations, one as a function of the redshift $z$ and
one  as a function of $\Omega_{DE}$. Although the fitting
parameters can be expressed as a function of the potential ones,
one can equivalently consider them as free parameters suitably
chosen to describe an arbitrary evolution. Due to the relative
large number of free parameters, such parametrizations are very
accurate up to present epoch. Thus one can use them in order to
describe various quintom evolution sub-classes, including
quintessence-like or phantom-like cases, realization of the
$-1$-crossing, non-monotonic $w(z)$ evolution etc.

The above analysis shows that a two-field (one canonical and one
phantom) quintom scenario with power-law potentials can offer a
good description of dark energy evolution, in agreement with
observations \cite{observ}. Moreover, the fact that power-law
potentials can be justified through supersymmetric considerations
\cite{Binetruy:1998rz} is an additional advantage. However, as it
is usual in models where phantom fields are present, the quantum
behavior of the examined scenario needs some caution, since the
discussion about the construction of quantum field theory of
phantoms is still open in the literature. For instance in
\cite{Cline:2003gs} the authors reveal the causality and stability
problems and the possible spontaneous breakdown of the vacuum into
phantoms and conventional particles, but on the other hand in
\cite{quantumphantom0} the phantom fields arise as an effective
description, consistently with the basic requirements of quantum
field theory. The comparison with observations and a robust
theoretical justification should be the decisive tests for the
present model.


\begin{thebibliography}{99}

\bibitem{observ}
A.G. Riess {\it et al.} [Supernova Search Team Collaboration],
Astron. J. {\bf 116}, 1009 (1998);
 S.
Perlmutter {\it{et al.}} [Supernova Cosmology Project
Collaboration], Astrophys. J. {\bf 517}, 565 (1999); A. Melchiorri
{\it et al.}, Astrophys. J. Lett. {\bf 536}, L63 (2000);
 D. N. Spergel {\it{et al.}}, Astrophys.
J. Suppl. {\bf 148}, 175 (2003); Bridle, O. Lahab, J. P. Ostriker
and P. J. Steinhardt, Science {\bf 299}, 1532 (2003); J. L. Tonry
{\it{et al.}}, Astrophys. J. {\bf 594}, 1 (2003); S. W. Allen,
{\it{et al.}}, Mon. Not. Roy. Astron. Soc. {\bf 353}, 457 (2004).

\bibitem{ordishov}
 P. Bin\'{e}truy, C. Deffayet, D. Langlois, Nucl. Phys. B {\bf565}, 269 (2000);
G.R. Dvali, G. Gabadadze, M. Porrati, Phys. Lett. B {\bf485}, 208
(2000); S. Capozziello, Int. J. Mod. Phys. D {\bf11}, 483 (2002);
 S.Nojiri
and S.~D.~Odintsov, Phys. Rev. D {\bf{68}}, 123512 (2003);
  A.~Lue and G.~Starkman,
  Phys.\ Rev.\  D {\bf 67}, 064002 (2003);
  P.~S.~Apostolopoulos, N.~Brouzakis, E.~N.~Saridakis and N.~Tetradis,
  Phys.\ Rev.\  D {\bf 72}, 044013 (2005);
  G.~Calcagni, S.~Tsujikawa and M.~Sami,
  Class.\ Quant.\ Grav.\  {\bf 22}, 3977 (2005);
S.Nojiri and S.~D.~Odintsov, Int. J. Geom. Meth. Mod. Phys.
{\bf{4}}, 115 (2007);
  R.~P.~Woodard,
  Lect.\ Notes Phys.\  {\bf 720}, 403 (2007);
  F.~K.~Diakonos and E.~N.~Saridakis,
 JCAP {\bf 0902}, 030 (2009).



\bibitem{quint0}
 P.~J.~E.~Peebles and  B.~Ratra, \apj {\bf 325}, L17 (1988);
 B.~Ratra and P.~J.~E.~Peebles, Phys. Rev. D {\bf 37}, 3406 (1988).

\bibitem{quint}
C.~Wetterich, Nucl.\ Phys.\ B {\bf 302}, 668 (1988); M. S. Turner
and M. White, Phys. Rev. D {\bf{56}}, 4439 (1997); R. R. Caldwell,
R. Dave and P. J. Steinhardt, Phys. Rev. Lett. {\bf{80}}, 1582
(1998); Z.~K.~Guo, N.~Ohta and Y.~Z.~Zhang, Mod.\ Phys.\ Lett.\ A
{\bf 22}, 883 (2007);
  O.~Hrycyna and M.~Szydlowski,
  Phys.\ Rev.\  D {\bf 76}, 123510 (2007);
  S.~Dutta, E.~N.~Saridakis, R.~J.~Scherrer, arXiv:0903.3412
[astro-ph.CO].



\bibitem{quint01}
I.~Zlatev, L.~Wang, and P.~J.~Steinhardt, Phys. Rev. Lett. {\bf
82}, 896 (1999);
 A.~R.~Liddle and R.~J.~Scherrer, Phys. Rev. D {\bf 59}, 023509
 (1999);
 P.~J.~Steinhardt, L.~Wang, and I.~Zlatev, Phys. Rev. D {\bf 59}, 123504 (1999).




\bibitem{phant} R. R. Caldwell, Phys.
Lett. B {\bf{545}}, 23 (2002); R.~R.~Caldwell, M.~Kamionkowski and
N.~N.~Weinberg, Phys. Rev. Lett. {\bf 91}, 071301 (2003);
 V. K.
Onemli and R. P. Woodard, Phys.\ Rev.\ D {\bf 70}, 107301 (2004);
  P.~F.~Gonzalez-Diaz and C.~L.~Siguenza,
  Nucl.\ Phys.\  B {\bf 697}, 363 (2004);
  M.~Sami, A.~Toporensky, P.~V.~Tretjakov and S.~Tsujikawa,
  Phys.\ Lett.\  B {\bf 619}, 193 (2005);
  A.~Vikman,
  Phys.\ Rev.\  D {\bf 71}, 023515 (2005);
  Z.~K.~Guo, R.~G.~Cai and Y.~Z.~Zhang,
  JCAP {\bf 0505}, 002 (2005);
S. Nojiri and S. D. Odintsov, Phys. Rev. D  {\bf{72}}, 023003
(2005);  H. Garcia-Compean, G. Garcia-Jimenez,  O. Obregon, and C.
Ramirez, JCAP {\bf 0807}, 016 (2008);
  E.~N.~Saridakis,
  arXiv:0811.1333 [hep-th];
  M.~Szydlowski and O.~Hrycyna,
  JCAP {\bf 0901}, 039 (2009);
  X.~m.~Chen, Y.~Gong and E.~N.~Saridakis,
  arXiv:0812.1117 [gr-qc].

\bibitem{quintom}
B.~Feng, X.~L.~Wang and X.~M.~Zhang, Phys.\ Lett.\  B {\bf 607},
35 (2005);
Z. K. Guo, {\it{et al.}}, Phys. Lett. B {\bf 608}, 177 (2005);
  H.~Wei, R.~G.~Cai and D.~F.~Zeng,
  Class.\ Quant.\ Grav.\  {\bf 22}, 3189 (2005);
M.-Z Li, B. Feng, X.-M Zhang, JCAP,  {\bf0512}, 002 (2005);
  P.~x.~Wu and H.~w.~Yu,
  Int.\ J.\ Mod.\ Phys.\  D {\bf 14}, 1873 (2005);
  R.~Lazkoz and G.~Leon,
  Phys.\ Lett.\  B {\bf 638}, 303 (2006);
  Y.~f.~Cai, H.~Li, Y.~S.~Piao and X.~m.~Zhang,
  Phys.\ Lett.\  B {\bf 646}, 141 (2007);
 W.
Zhao and Y. Zhang, Phys. Rev. D {\bf73}, 123509 (2006);
 B. Feng, M.
Li, Y.-S. Piao and X. Zhang, Phys. Lett. B {\bf 634}, 101 (2006);
  M.~R.~Setare and E.~N.~Saridakis,
  Phys.\ Lett.\  B {\bf 668}, 177 (2008);
  M.~R.~Setare and E.~N.~Saridakis,
  JCAP {\bf 0809}, 026 (2008);
  M.~R.~Setare and E.~N.~Saridakis,
  Phys.\ Lett.\  B {\bf 671}, 331 (2009).

\bibitem{c14}
 P. S. Apostolopoulos, and N. Tetradis, Phys. Rev. D {\bf
74}, 064021 (2006); H.-S. Zhang, and Z.-H. Zhu, Phys. Rev. D {\bf
75}, 023510 (2007);
  Y.~Gong and A.~Wang,
  Phys.\ Lett.\  B {\bf 652}, 63 (2007);
  E.~N.~Saridakis, P.~F.~Gonzalez-Diaz and C.~L.~Siguenza,
  arXiv:0901.1213 [astro-ph].



\bibitem{Scherrer:2007pu}
  R.~J.~Scherrer and A.~A.~Sen,
  Phys.\ Rev.\  D {\bf 77}, 083515 (2008);
  R.~J.~Scherrer and A.~A.~Sen,
  Phys.\ Rev.\  D {\bf 78}, 067303 (2008);
  M.~R.~Setare and E.~N.~Saridakis,
  Phys.\ Rev.\  D {\bf 79}, 043005 (2009).


\bibitem{Binetruy:1998rz}
  P.~Binetruy,
  Phys.\ Rev.\  D {\bf 60}, 063502 (1999);
  A.~Masiero, M.~Pietroni and F.~Rosati,
  Phys.\ Rev.\  D {\bf 61}, 023504 (2000).


\bibitem{Kneller:2003xg}
  J.~P.~Kneller and L.~E.~Strigari,
  Phys.\ Rev.\  D {\bf 68}, 083517 (2003);
  L.~R.~W.~Abramo and F.~Finelli,
  Phys.\ Lett.\  B {\bf 575}, 165 (2003);
    X.~Zhang,
  Mod.\ Phys.\ Lett.\  A {\bf 20}, 2575 (2005);
  M.~Yashar, B.~Bozek, A.~Abrahamse, A.~J.~Albrecht and M.~Barnard,
  arXiv:0811.2253 [astro-ph].


\bibitem{Watson:2003kk}
  C.~R.~Watson and R.~J.~Scherrer,
  Phys.\ Rev.\  D {\bf 68}, 123524 (2003).


\bibitem{Saridakis:2009pj}
  E.~N.~Saridakis,
  arXiv:0902.3978 [gr-qc].



\bibitem{Szydlowski:2006ay}
G. Efstathiou, Mon. Not. R. Astron. Soc. {\bf 310}, 842 (1999);
  A.~R.~Cooray and D.~Huterer,
  Astrophys.\ J.\  {\bf 513}, L95 (1999);
    M.~Chevallier and D.~Polarski,
  Int.\ J.\ Mod.\ Phys.\  D {\bf 10}, 213 (2001);
  E.~V.~Linder,
  Phys.\ Rev.\ Lett.\  {\bf 90}, 091301 (2003).
  E.~Di Pietro and J.~F.~Claeskens,
  Mon.\ Not.\ Roy.\ Astron.\ Soc.\  {\bf 341}, 1299 (2003);
  M.~Szydlowski, A.~Kurek and A.~Krawiec,
  Phys.\ Lett.\  B {\bf 642}, 171 (2006);
  E.~J.~Copeland, M.~Sami and S.~Tsujikawa,
  Int.\ J.\ Mod.\ Phys.\  D {\bf 15}, 1753 (2006);
  H.~Wei and S.~N.~Zhang,
  Phys.\ Lett.\  B {\bf 644}, 7 (2007).


    \bibitem{Cline:2003gs}
    J.~M.~Cline, S.~Jeon and G.~D.~Moore,
    Phys.\ Rev.\  D {\bf 70}, 043543 (2004).

\bibitem{quantumphantom0}
  S.~Nojiri and S.~D.~Odintsov,
  Phys.\ Lett.\  B {\bf 562}, 147 (2003);
  S.~Nojiri and S.~D.~Odintsov,
  Phys.\ Lett.\  B {\bf 571}, 1 (2003).










\end{thebibliography}
\end{document}